\documentclass{emulateapj}

\slugcomment{}
\usepackage{graphicx}
\usepackage{longtable}

\shorttitle{{\it Herschel} SPIRE Sees Dust Emission of Galaxies at $z=4$}
\shortauthors{Lee et al. }
\begin{document}
\def\hh{\, h^{-1}}
\newcommand{\ie}{$i.e.$,}
\newcommand{\wth}{$w(\theta)$}
\newcommand{\mpc}{Mpc}
\newcommand{\xir}{$\xi(r)$}
\newcommand{\Lya}{Ly$\alpha$}
\newcommand{\Lyb}{Lyman~$\beta$}
\newcommand{\Hb}{H$\beta$}
\newcommand{\msun}{M$_{\odot}$}
\newcommand{\hmsun}{$h^{-1}$M$_{\odot}$}
\newcommand{\sfr}{M$_{\odot}$ \text{yr}$^{-1}$}
\newcommand{\dnsty}{$h^{-3}$Mpc$^3$}
\newcommand{\za}{$z_{\rm abs}$}
\newcommand{\ze}{$z_{\rm em}$}
\newcommand{\cmtwo}{cm$^{-2}$}
\newcommand{\nhi}{$N$(H$^0$)}
\newcommand{\degpoint}{\mbox{$^\circ\mskip-7.0mu.\,$}}
\newcommand{\halpha}{\mbox{H$\alpha$}}
\newcommand{\hbeta}{\mbox{H$\beta$}}
\newcommand{\hgamma}{\mbox{H$\gamma$}}
\newcommand{\kms}{\,km~s$^{-1}$}      
\newcommand{\minpoint}{\mbox{$'\mskip-4.7mu.\mskip0.8mu$}}
\newcommand{\mv}{\mbox{$m_{_V}$}}
\newcommand{\Mv}{\mbox{$M_{_V}$}}
\newcommand{\peryr}{\mbox{$\>\rm yr^{-1}$}}
\newcommand{\secpoint}{\mbox{$''\mskip-7.6mu.\,$}}
\newcommand{\sqdeg}{\mbox{${\rm deg}^2$}}
\newcommand{\squig}{\sim\!\!}
\newcommand{\subsun}{\mbox{$_{\twelvesy\odot}$}}
\newcommand{\et}{{\it et al.}~}
\newcommand{\er}[2]{$_{-#1}^{+#2}$}
\def\h50{\, h_{50}^{-1}}
\def\hbl{km~s$^{-1}$~Mpc$^{-1}$}
\def\ltsima{$\; \buildrel < \over \sim \;$}
\def\simlt{\lower.5ex\hbox{\ltsima}}
\def\gtsima{$\; \buildrel > \over \sim \;$}
\def\simgt{\lower.5ex\hbox{\gtsima}} 
\def\arcs{$''~$}
\def\arcm{$'~$}
\newcommand{\wu}{$U$}
\newcommand{\wb}{$B_{435}$}
\newcommand{\wv}{$V_{606}$}
\newcommand{\wi}{$i_{775}$}
\newcommand{\wz}{$z_{850}$}
\newcommand{\hmpc}{$h^{-1}$Mpc}
\newcommand{\lm}{$L$--$M$}
\newcommand{\ws}{$\mathcal{S}$}
\newcommand{\wm}{$\mathcal{M}$}
\newcommand{\sm}{$\mathcal{S}$-$\mathcal{M}$}
\newcommand{\medianLM}{$\tilde{\mathcal{L}}(M)$}
\newcommand{\mf}{$\phi_\mathcal{M}$}


\title{{\it Herschel} Detection of Dust Emission from UV-Luminous Star-Forming Galaxies at $3.3\simlt z \simlt 4.3$ }

\author{Kyoung-Soo Lee\altaffilmark{1}, Stacey Alberts\altaffilmark{2}, David Atlee\altaffilmark{3}, Arjun Dey\altaffilmark{3}, Alexandra Pope\altaffilmark{2}, \\Buell  T. Jannuzi\altaffilmark{3,4}, Naveen Reddy\altaffilmark{5}, Michael J.~I.~Brown\altaffilmark{6}}
\altaffiltext{1}{Department of Physics, Purdue University, 525 Northwestern Avenue, West Lafayette, IN 47907}
\altaffiltext{2}{Department of Astronomy, University of Massachusetts, 710 North Pleasant Street, Amherst, MA 01003}
\altaffiltext{3}{National Optical Astronomy Observatory, Tucson, AZ 85726}
\altaffiltext{4}{Steward Observatory, University of Arizona, Tucson, AZ 85721}
\altaffiltext{5}{Department of Physics and Astronomy, University of California, Riverside, 900 University Avenue, Riverside, CA 92521, USA}
\altaffiltext{6}{Monash University, Clayton, Victoria 3800, Australia}

\begin{abstract}
We report the {\it Herschel} SPIRE detection of dust emission arising from UV-luminous ($L\gtrsim L^*$) star-forming galaxies at $3.3\simlt z\simlt 4.3$. Our sample of 1,913 Lyman Break Galaxy (LBG) candidates is selected over an area of 5.3 deg$^2$ in the Bo\"otes Field of the NOAO Deep Wide-Field Survey. This is one of the largest samples of UV-luminous galaxies at this epoch and enables an investigation of the bright end of the galaxy luminosity function. We divide our sample into three luminosity bins and stack the {\it Herschel} SPIRE data  to measure the average spectral energy distribution (SED) of LBGs at far-infrared (FIR) wavelengths. We find that these galaxies have average IR luminosities of $(3-5)\times 10^{11}L_\odot$ and $60-70$\% of their star-formation obscured by dust. The FIR SEDs peak at $\lambda_{\rm rest}\simgt 100\mu m$, suggesting dust temperatures ($T_d=27-30~K$) significantly colder than that of local galaxies of comparable IR luminosities.  The observed IR-to-UV luminosity ratio (${\rm IRX}\equiv L_{\rm IR}/L_{\rm UV}$) is low ($\approx 3-4$) compared with that observed for $z\approx2$ LBGs (${\rm IRX_{z\sim2}}\approx 7.1\pm1.1)$. The correlation between the slope of the UV continuum and IRX for galaxies in the two lower luminosity bins suggests dust properties similar to those of local starburst galaxies. However, the galaxies in the highest luminosity bin appear to deviate from the local relation, suggesting that their dust properties may differ from those of their lower-luminosity and low-redshift counterparts. We speculate that the most UV luminous galaxies at this epoch are being observed in a short-lived and young evolutionary phase. 

\end{abstract}
  

\section{Introduction}

Measurements of the spectral energy distribution (SED) of high-redshift galaxies can provide insight into their constituents (i.e., stars, gas, dust, AGN), star-formation rates (SFRs), and formation history. While high-redshift ($z\simgt 2$) star-forming galaxies have been identified in large numbers using rest-frame UV color selection techniques \citep[e.g.,][]{steidel96,steidel03,adelberger04,bouwens10b},  their faintness across the electromagnetic spectrum has limited our understanding of their multiwavelength properties. In particular, UV emission is easily extinguished by dust and re-radiated at far-infrared (FIR) wavelengths. Without direct measurements of the FIR emission, the amount of dust extinction and the intrinsic SFRs remain uncertain. Thus far, our understanding of the SFRs largely rests on the assumption that the UV-selected high-$z$ galaxies are similar in their dust properties to local starburst galaxies, and specifically that the dust extinction in the UV (as measured by the slope, $\beta$, of the UV continuum) can be used to predict the infrared luminosity ($L_{\rm IR}$) due to dust emission \citep[e.g.,][]{meurer99,calzetti00,reddy08,finkelstein09}. 

The  {\it Spitzer} and {\it Herschel}\footnote[7]{{\it Herschel} is an ESA space observatory with science instruments provided by European-led Principal Investigator consortia and with important participation from NASA}  Space Telescopes have enabled direct measurements of the mid- and FIR emission from distant galaxies and thus a direct test of this assumption. While these telescopes still lack the sensitivity to individually detect any but the most IR-luminous  high-$z$ galaxies, stacking analyses of $z\sim2-3$ UV-selected galaxy samples have resulted in useful constraints \citep{rigopoulou10,magdis10b,reddy12_herschel}. Several studies of galaxies at $z\sim2-3$ \citep[][]{reddy06a,daddi07a,magdis10a,reddy10,reddy12_herschel} have demonstrated that the ratio of the average UV continuum slope to the average far-infrared luminosity is similar to that in local starburst galaxies. There are exceptions: most notably, there is evidence that the local $\beta$-$L_{\rm IR}$ relation may not apply to very dusty, UV-faint galaxies, or to extremely young galaxies \citep[e.g.,][]{goldader02, reddy10}. Also, the studies thus far are based primarily on galaxy samples with luminosity ranges that do not include large numbers of the most luminous galaxies, and the dependence of the $\beta$-$L_{\rm IR}$ relation on luminosity is not well understood at high-$z$. 

In this paper, we extend this test to higher-redshift and higher-luminosity UV-selected galaxies. We investigate the far-IR properties of a very large sample of $3\simlt z \simlt 4.5$ galaxies.  We stack the {\it Herschel} Spectral and Photometric Imaging Receiver \citep[SPIRE:][]{spire} observations of this sample and present the first (stacked) detections of the far-IR emission of UV-selected galaxies at this epoch. We use a  $\Lambda$CDM cosmology with $(\Omega, \Omega_\Lambda, \sigma_8, h_{100})=(0.28, 0.72, 0.9, 0.72)$.  

\section{Data and Galaxy Sample}

The datasets used in our analyses are:  optical data ($B_WRI$) in the Bo\"otes field of the NOAO Deep Wide-Field Survey  \citep[][]{jannuzid99}; near-infrared $JHK_S$ data taken with the NEWFIRM camera (Gonzalez et al., in prep); mid- and far-infrared (3.6--160$\mu m$) data from the {\it Spitzer} Infrared Array Camera (IRAC) and the Multiband Imaging Photometer (MIPS) obtained by the {\it Spitzer} Deep Wide-Field Survey \citep{ashby09} and the MIPS AGN and Galaxy Evolution Survey (Jannuzi et al., in~prep); and far-IR (250--500$\mu m$) {\it Herschel} SPIRE observations obtained by the {\it Herschel} Multi-tiered Extragalactic Survey \citep[HerMES:][]{oliver12}.   We reduced and mosaiced the SPIRE data using the Herschel Interactive Processing Environment \citep[HIPE;][]{hipe}, removing striping, astrometry offsets, and glitches missed by the standard pipeline (Alberts et al., in prep). The SPIRE maps include a deep 2 deg$^2$ region surrounded by a shallower outer region. The $1\sigma$ depths at 250, 350 and 500$\mu m$ are 3, 2, 3 mJy in the inner region and 5, 4, 5 mJy in the outer region respectively. 

We use broad-band ($B_WRI$) color selection to identify sources that exhibit a continuum break between the $B_W$ and $R$ bands \citep{lee11}. This technique, pioneered by \citet{steidelh93}, successfully identifies $z\sim4$ UV-emitting star-forming galaxies (also known as Lyman Break Galaxies, or LBGs) which exhibit a strong break in their rest-frame UV spectra at 912\AA\ (the Lyman limit) and additional  flux attenuation due to the intervening Lyman alpha forest at $912-1216$\AA. Spectroscopic redshifts of $\approx$5\% of the sample \citep[][Lee et al. 2012, in prep]{lee11,kochanek12} have confirmed a broad redshift distribution ($3\simlt z \simlt 4.5$, $\bar{z}\approx3.7; \sigma_z\approx 0.4$), and a contamination rate of $8-23\%$ (the limits representing the optimistic/pessimistic scenario based on the nature of the ambiguous spectra). Our final sample consists of 1,913 $L_{\rm{UV}}\gtrsim L^*$  LBG candidates selected over a contiguous 5.3 deg$^2$ region of the Bo\"otes field \citep[see][]{lee11}. 

We divide the sample into three bins according to the $I$-band magnitudes as $I_{\rm{AB}}=[21.3,23.7], [23.7,24.3],[24.3,24.7]$. For reference, the characteristic luminosity $L^*$ at $\bar{z}=3.7$ corresponds to $I_{\rm{AB}}=24.7$ \citep{bouwens07}. The magnitude bins correspond to overlapping luminosity ranges due to the finite width of the redshift distribution. By using the observed magnitude and redshift distributions of the sample, we find that the magnitude bins correspond to roughly Gaussian bins of $L_{\rm UV}/L^*$, with widths at half-maxima of [$>$2.3], [1.4,2.5], and [0.9,1.5], respectively. We refer to these bins as the ``high-'', ``intermediate-'', and ``low-luminosity'' sample. The majority (97\%) of galaxies in the high-luminosity bin  are $I>22$. The median UV luminosities of our samples are presented in Table~\ref{tbl_1}. 

\begin{deluxetable*}{lccccccc}
\tabletypesize{\scriptsize}
\tablewidth{0pc}
\tablecaption{Summary of UV-to-FIR properties of Star-Forming Galaxies at $z\sim3.7$}
\tablehead{
\colhead{Sample} &
\colhead{N$_{\rm{gal}}$\tablenotemark{a}} &
\colhead{$\log[L_{\rm{UV}}]$\tablenotemark{b} } &
\colhead{$\log[L_{\rm{IR}}]$\tablenotemark{b} } &
\colhead{IRX}&
\colhead{$\beta$} & 
\colhead{SFR$_{\rm{UV}}$\tablenotemark{c}} &
\colhead{SFR$_{\rm{IR}}$\tablenotemark{c}}}
\startdata
$I_{\rm{AB}}$=[21.3,23.7] & 325/293 & $11.21\pm0.06$ & $(11.65-11.74)$ & $2.7-3.3$ & $-1.24\pm0.08$ & $49\pm7$ & $77-95$ \\
$I_{\rm{AB}}$=[23.7,24.3] & 784/686 & $10.96\pm0.06$& $(11.48-11.57)$ & $3.4-4.1$ & $-1.65\pm0.11$& $28\pm4$ & $52-64$\\
$I_{\rm{AB}}$=[24.3,25.0] & 794/694 & $10.76\pm0.06$ & $<(11.50-11.57; 3\sigma)$ & $<(5.5-6.4)$ & $-1.75\pm0.15$ & $18\pm3$ & $<(54-63)$ \\
\enddata
\tablenotetext{a}{The number of LBGs and the number of LBGs included in the {\it Herschel} stacking}
\tablenotetext{b}{In units of $L_\odot$}
\tablenotetext{c}{In units of $M_\odot{\rm yr}^{-1}$. The \citet{kennicutt98} calibration is used  assuming the Salpeter IMF ($0.1-100M_\odot$), solar metallicity}
\label{tbl_1}
\end{deluxetable*}

\section{Data Stacking}

We measured the average optical-to-mid-IR SED of the galaxy samples by using photometry of stacked images in the individual optical, near-IR and IRAC bands. We first convolved all images in a given band to a common point-spread function (PSF), then extracted a cutout centered on each source, subtracted a local sky background, and produced a composite image by taking the median flux over all sources at each pixel position \citep[for further details, see][]{lee11}.

The {\it Spitzer} MIPS and {\it Herschel} data both suffer from large beams and spatial variations in the sensitivity.  For the  MIPS data, we used a variance-weighted median to account for small-scale variations in the S/N of the images as a function of position. The stacked data only yielded upper limits. The $3\sigma$  limits, measured in apertures of diameter $6\arcsec$, $18\arcsec$, $38\arcsec$ (matched to the full-width-at-half-maximum of the PSFs in the $24\mu m$, $70\mu m$ and $160\mu m$ images, respectively) are presented in Table~\ref{fluxtable}. 

The HerMES SPIRE data only cover a portion of our field, and sample 1673 LBGs (i.e., 85\%). Thirty five (2\%) LBGs have a 250$\mu m$ detection ($\ge5\sigma$) within 8\arcsec; 33 of these have more than one optical counterpart in the beam. Of these, only two sources exhibit increasing flux densities towards longer wavelengths, as would be expected for $z>3.5$ galaxies (sampling the Wien side of the dust emission). However, both  have another optical counterpart other than the LBG candidate within the beam, making it uncertain whether  the far-IR emission is uncontaminated. We therefore assumed that the detected sources are contaminated by unresolved interlopers, and excluded all 35 sources from the stacking. We stacked the  SPIRE data by calculating the variance-weighted average flux.  The measurement significance  was determined using a bootstrap resampling technique \citep[e.g.,][]{numrec}. We measured the flux using 10,000 random samples of the source list (replacing chosen sources each time) and measured the RMS from the resulting flux distribution. Our results are presented in Table~\ref{fluxtable}. For completeness, we repeated our analyses with and without the 250$\mu m$ detected sources; the differences are within the $1\sigma$ uncertainties.

High-redshift submillimeter galaxies are known to be strongly clustered \citep{blain04,hickox12} and are often found in close proximity to LBGs \citep[][]{chapman09}. 
The large SPIRE beam could be contaminated by such sources.  In order to estimate the degree of contamination, we used the substantially deeper 24$\mu m$ MIPS data and measured  photometry in large apertures matched to the SPIRE beamsize. We carried out this measurement at both on-source (LBGs) and random sky positions. We found that (1) there is no significant difference between the on-source and sky distributions, and (2) the distributions are similar at both the 250$\mu m$ and 500$\mu m$ beamsizes. This suggests that the contribution of clustered sources to the flux measured for the LBGs is small. Therefore, our stacked fluxes likely represent the FIR emission from a typical $z\sim3.7$ LBG.

\begin{table}
\caption{Flux Densities in the Stacked MIPS and {\it Herschel} Images\label{tbl_2}}
\begin{tabular}{lccc}
\hline
\hline
 & Sample 1 & Sample 2 & Sample 3\\
 & $I_{\rm{AB}}$=[21.3,23.7]  &$I_{\rm{AB}}$=[23.7,24.3]  & $I_{\rm{AB}}$=[24.3,24.7] \\
\hline
$24\mu m$ &$<0.016$ & $<0.011$ & $<0.011$ \\
$70\mu m$ & $<1.38 $ & $<0.90$ & $<0.90$ \\
$160\mu m$ & $<13.2$ & $<9.54$ & $<9.92$ \\
$250\mu m$ & $<1.05$ & $<0.84$ & $<0.72$ \\
                        & $<1.11$ & $<0.90$ & $<0.78$ \\
$350\mu m$ &  $0.65\pm0.39$ & $0.44\pm0.29$ & $<0.78$ \\
                        & $1.01\pm0.43$ & $0.85\pm0.33$ & $<0.87$ \\
$500\mu m$ & $1.45\pm0.47$ & $1.06\pm 0.31$ & $< 0.90$\\
                        & $1.69\pm0.48$ & $1.37\pm0.33$ & $<0.96$\\                      
\hline
\end{tabular}
\tablecomments{The flux densities are in units of mJy and the upper limits are 3$\sigma$ in flux. 
For the {\it Herschel} data, the first and second row at each observing wavelength represent the stacked fluxes excluding and including sources with a $>5\sigma$ detection at 250$\mu m$, respectively}. 
\label{fluxtable}
\end{table}

\section{Results and Discussion}

\subsection{The Panchromatic View of UV-Luminous Star-Forming Galaxies at $z\sim3.7$}

\begin{figure*}[t]
\epsscale{1.}
\plotone{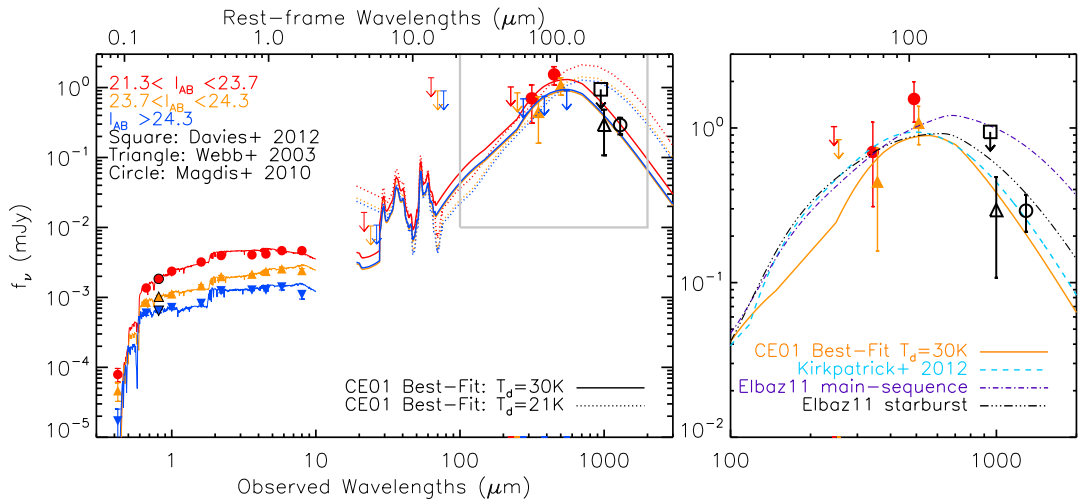}
\caption[stacked_sed]
{
{\it Left:} The median SEDs ($0.3-500\mu m$) of the three samples are shown with the best-fit stellar population models and local IR templates of \citet{chary01}.  The {\it Herschel} points are slightly offset (in wavelength)  for clarity; the intermediate-luminosity bin is positioned at the correct wavelengths.  The peak of the dust emission is not well constrained by the current data and we therefore show the acceptable models with both the lowest/highest dust temperatures ($T_d=21/30$K). The latter model is favored by the sub-mm/mm measurements of similarly-selected LBGs \citep{webb03,magdis10a,davies12}.  {\it Right:} The same far-infrared measurements  are compared with three {\it Herschel}-based templates of \citet{elbaz11} and \citet{kirkpatrick12}. All templates are normalized to match the intermediate-luminosity points (filled triangles).
}
\label{stacked_sed}
\end{figure*}

Figure~\ref{stacked_sed} shows the stacked SEDs of the LBGs in the three magnitude bins. At $\lambda_{\rm{rest}}<10\mu m$, the SEDs show similar shapes, with two noticeable differences. First, the UV continuum slope $\beta$ ($f_\lambda \propto \lambda^\beta$ measured at  $\lambda_{\rm rest}$=1300-3000\AA) increases with luminosity from $\beta=-1.75\pm0.15$ (low-luminosity bin) to $-1.24\pm 0.08$ (high-luminosity bin). 
Shifting the median redshift changes the level of intergalactic attenuation and but  does not impact the observed colors significantly (e.g., $\Delta z=0.1$ would result in   $\Delta\beta \simlt 0.05$). Second, the break observed between $J$ and $K_S$ decreases in strength with increasing luminosity. If this is the Balmer  break, the trend suggests that galaxies in the high-luminosity bin are, on average, younger than their lower luminosity counterparts \citep[e.g.,][]{bc93}.

In the far-IR, only the intermediate- and high-luminosity bins show detections in the 350$\mu m$ and 500$\mu m$ SPIRE bands. To measure the total IR luminosities, we modeled the SEDs using the templates of \citet[][hereafter CE01]{chary01}. The CE01 templates are designed to match the SEDs of present-day IR-luminous galaxies and to reproduce the observed local correlation between dust temperature  ($T_{\rm{dust}}$) and IR luminosity. 
If we only use CE01 templates that correspond to the observed IR luminosity ranges, the resulting fits are poor ($\chi_r^2\ge2.3$; 90\% confidence). This is primarily because the observed SEDs of our samples peak at longer wavelengths than those of local galaxies of comparable IR luminosities. The longer peak wavelength observed for our galaxies may suggest that galaxies at $z\sim3.7$ have colder dust than local galaxies at similar IR luminosities. Similar results have been found for various types of high-redshift ($z>1.5$) galaxies \citep[][]{pope06, dannerbauer10, muzzin10,elbaz11,kirkpatrick12}. 

To account for possible redshift evolution of IR SEDs, we fit all the CE01 templates leaving the normalization as a free parameter, effectively eliminating the local $T_{\rm{dust}}$-$L_{\rm{IR}}$ correlation from the fitting procedure.  The $3\sigma$ upper limits were included in the $\chi^2$ calculation only when the model exceeds the limit. For the low-luminosity bin with no {\it Herschel} detection, we determined the upper limit on the IR luminosity using the best-fit templates of the two other bins. The total IR luminosities ($L_{\rm IR}$: integrated between $8-1000\mu m$) are computed by integrating the scaled best-fit CE01 template. Typically, $\chi_r^2\sim0.1$ for the best-fit models.  

We find $L_{\rm IR}$ values of $(4.5-5.5) \times 10^{11}L_\odot$, $(3.0-3.7)\times 10^{11}L_\odot$, and $<(3.2-3.7)\times 10^{11}L_\odot$, in the order of decreasing UV luminosities (Table~\ref{tbl_1}).  Despite the uncertainties, a clear trend emerges: more UV-luminous galaxies are also more IR-luminous. The main source of uncertainty is the peak location of dust emission (and therefore $T_d$). The lower limit corresponds to $T_d\approx 30$K (modeled as a grey-body function $f_\nu\propto B_\nu(T_d,\nu)\nu^\beta$ where $B_\nu$ is the Planck function, $\beta$=1.5 is the dust emissivity). The upper limit is set by the coldest CE01 template ($T_d\approx21$K). 

As all SPIRE bands lie at shorter wavelengths than the FIR peak at $z\sim3.7$, we are unable to constrain the SED shape from our data alone. 
 However, it is instructive to consider the existing measurements at longer wavelengths on other UV-selected LBG samples at $z \gtrsim 3$. At $z\sim3$ \citet{webb03} reported a $1.5\sigma$ detection at 850$\mu m$ (using SCUBA) for $L\gtrsim L^*$ LBGs, while \citet{magdis10a} found a $3.7\sigma$ detection at 1.1mm (using AzTEC) of IRAC-detected LBGs. At $z\sim5$, \citet{davies12} determined a $2\sigma$ upper limit of 1.6 mJy at 1.2mm (using MAMBO). In Figure~\ref{stacked_sed}, we show these points after correcting the wavelengths and flux densities to $z=3.7$. Even though the different selection methods and redshift ranges of these studies make direct comparisons difficult, the measurements still provide a useful constraint on the SED, preferring a peak wavelength near $\approx500\mu m$. The inferred dust temperature is $T_d=(27-30)$ K. When compared to recent {\it Herschel}-based templates, the combined FIR constraints are in good agreement with the $z\sim2$ star-forming template of \citet{kirkpatrick12} and the starburst template of \citet{elbaz11}. The main-sequence Elbaz template is ruled out ($>2.5\sigma$) as it predicts $3-4\times$ higher flux densities in the sub-mm/mm wavelengths \citep[see][for further discussion]{kirkpatrick12}. Further constraints can be easily obtained by ALMA. Our best-fit model for the $1.5L^*-2.5L^*$ LBGs  predicts the flux densities of 0.83~(0.51) mJy in the $450~(870) \mu m$, which, according to the ALMA Sensitivity Calculator, can be detected ($5\sigma$) in 2.1~hr (5.6~min).

\subsection{Dust-Obscured Star Formation}
Having measured the IR luminosity of the LBG samples, we now compare the relative energy budget of star-formation in the UV and infrared. For each bin, we estimate  the luminosity ratio IRX$\equiv L_{\rm IR}/L_{\rm UV}$ assuming two dust temperatures,  $T_d=30$K and 21K (see Table \ref{tbl_1}). In all luminosity bins, the IRX remains relatively low at ${\rm IRX}\approx 3-4$, implying $60-70$\% of the star formation in these LBGs is obscured by dust. The IRX is weaker than that observed at $z\sim2$ \citep[$\approx$80\%:][]{reddy12_herschel}, suggesting the buildup of dust with cosmic time as low-mass stars continue to evolve into the AGB phase \citep[e.g.,][]{bouwens09}. 

Next, we  investigate how dust extinction inferred from the UV compares with  dust emission in the IR.  \citet{meurer99} derived the scaling law between dust extinction and the UV slope $\beta$ based on the observed correlation for local starbursts. Such a correlation is explained if local starburst galaxies are governed by the \citet{calzetti00} extinction law, which gives a flatter attenuation curve than that found for the Large and Small Magellanic Clouds \cite[][]{fitzpatrick86,bouchet85}.

At $z\sim2$, \citet{reddy10,reddy12_herschel} found that most UV-selected star-forming galaxies follow the local IRX-$\beta$ relation, demonstrating that the dust properties and relative geometry between dust and stars within the $z\sim2$ star-forming galaxies are similar to those in local starbursts \citep[see also][]{nandra02,daddi07a, magdis10a}. However, there are also exceptions: several galaxies lie off this relation (Reddy et al 2006, 2012; Siana et al. 2008, 2009)
 and are interpreted as being very young ($<$100~Myr).

\begin{figure}[t]
\epsscale{1.}
\plotone{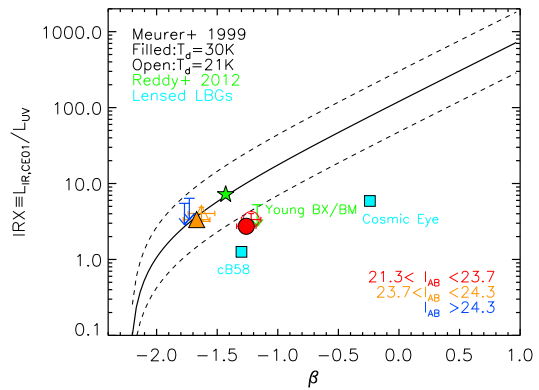}
\caption[IRX_beta]
{
The IRX values as a function of the UV slope $\beta$: The local Meurer relation, is shown with the 0.4~dex scatter. We show the locations of three subsamples at $z\sim3.7$ assuming  two different dust temperatures ($T_d=30$K, 20K in filled and open symbols, respectively).  Both estimates are slightly offset in $\beta$ for clarity. The locations of the $z\sim2$ galaxies, young ($<100$~Myr) $z\sim2$ galaxies \citep[][]{reddy12_herschel}, and two lensed LBGs \citep{siana08,siana09} are also shown. The galaxies in the two less-luminous bins follow the local relation while  the galaxies in the high-luminosity bin lies outside the nominal range the Meurer relation close to young $z\sim2$ galaxies. 
}
\label{IRX_beta}
\end{figure}

Figure~\ref{IRX_beta} shows the locations of the three $z\sim3.7$ LBG subsamples on the IRX-$\beta$ plane and the local relation of \citet{meurer99}\footnote{The correlation is scaled by 0.24~dex to match our definition of $L_{\rm{IR}}$ as IR luminosity integrated at $8-1000\mu m$ from that of \citet[][at $40-120 \mu m$]{meurer99} }. For comparison, we also show the locations of the $z\sim 2\  L_{\rm{UV}}\approx L^*$  galaxies of \citet{reddy12_herschel} and two lensed galaxies \citep[the Cosmic Eye and MS1512-cB58:][]{siana08,siana09}. The  galaxies in the intermediate-luminosity bin closely trace the local relation. The upper limits for our low-luminosity sample are also consistent with the local relation. Hence, our results suggest that  the dust properties in the $L^*-2.5L^*$ LBGs at $z\sim3.7$ are similar to those of local starbursts. 

The high-luminosity bin lies significantly (0.5 dex) below the Meurer relation, closer to the location of the young $z\sim2$  galaxies \citep{reddy12_herschel}. The observed offset suggests that the dust properties of the galaxies in the high-luminosity bin may be different from those of their lower-luminosity counterparts. Also, since the $\beta$ values for these galaxies cannot predict the $L_{\rm IR}$ using the local IRX-$\beta$ relation, it implies that SFRs derived from the UV using the standard relationship overpredict the true SFR in these systems. The impact of this result on the total cosmic SFR density is minimal. Nevertheless, the offset in the IRX-$\beta$ plane has interesting implications for the evolutionary state of these high-luminosity galaxies. 

The high-luminosity LBGs appear to have a weaker Balmer break than those in less-luminous bins, implying that they may have younger population ages. \citet{reddy10, reddy12_herschel} made similar observations that $z\sim2$ galaxies with ages $<100$~Myr have lower IRX values.  This offset suggests that the most UV-luminous galaxies may be undergoing a short evolutionary phase during which the dust covering fraction and/or the dust grain size distribution is different than that in their less luminous counterparts (Siana et al. 2009, Reddy et al. 2010).

The statistics from the existing spectroscopy supports the possibility that a large dust covering fraction may be suppressing the emergence of Ly$\alpha$ photons in the most UV-luminous galaxies. Of the 13 galaxies at $I_{\rm{AB}}=22.5-24.3$, none has strong Ly$\alpha$ emission (rest-frame equivalent width $EW_0\le$20\AA). In contrast, 12/30 galaxies at $I_{\rm{AB}}=24.3-25.0$ have $EW_0>20$\AA. While the spectroscopic confirmation of the fainter LBGs may be biased toward the stronger Ly$\alpha$ emitters, the lack of Ly$\alpha$ emission among the most UV-luminous galaxies is significant  \citep[see][]{stark10,kornei10}. 

The suggested luminosity-age relation we observe in the $z\sim3.7$ LBGs is not observed for $z\sim2$ galaxies as the median luminosity of their youngest (IRX-outlier) population is similar to that of the overall sample. However, the luminosity range of our high-luminosity sample is not well represented in the Reddy~et~al. sample. It is also possible that our results may have been influenced by sample contamination. Interlopers such as dusty low-redshift ($z<1$) galaxies or low-luminosity quasars  could simulate a smaller Balmer break and depressed FIR flux. Based on our DEIMOS spectroscopy, the contamination rate in the high-luminosity bin is $\approx$15\%, slightly higher than that of the full sample ($10-12$\%).  Even if the FIR flux is zero for the interlopers, the resulting UV flux correction to the IRX would be insufficient to explain the observed offset in Figure 2. If the low-redshift interlopers contribute disproportionately to the measured average IR flux, the true IRX offset of our high-luminosity LBGs from the Meurer relation would be even larger than observed. Spectroscopic confirmation of a larger number of candidates can place more stringent limits on the interloper populations, and also allow the stacking of confirmed LBGs.

\acknowledgments
KSL thanks Mark Dickinson and Adam Muzzin for stimulating discussions. We thank the referee for a careful reading of the manuscript and a helpful report. This work is based in part on observations made with the Spitzer Space Telescope, which is operated by the Jet Propulsion Laboratory, California Institute of Technology under a contract with NASA. We acknowledge the MAGES and NDWFS teams and NOAO/NSF support for this research.


\end{document}